\RequirePackage{fix-cm}
\documentclass[smallextended]{svjour3}       
\smartqed  
\usepackage{graphicx}

\newcommand{\bfm}[1]{\mbox{\boldmath{$#1$}}}

\begin{document}

\title{Stability of the Euler Resting N-Body Relative Equlilbria
}
\subtitle{}


\author{D.J. Scheeres 
}


\institute{D.J. Scheeres \at
              429 UCB, Smead Department of Aerospace Engineering Sciences, The University of Colorado Boulder \\
              Tel.: +1-720-544-1260\\
              \email{scheeres@colorado.edu}           
}

\date{Received: date / Accepted: date}

\maketitle

\begin{abstract}
The stability of a system of $N$ equal sized mutually gravitating spheres resting on each other in a straight line and rotating in inertial space is considered. This is a generalization of the ``Euler Resting'' configurations previously analyzed in the finite density 3 and 4 body problems.  Specific questions for the general case are how rapidly the system must spin for the configuration to stabilize, how rapidly it can spin before the components separate from each other, and how these results change as a function of $N$. This paper shows that the Euler Resting configuration can only be stable for up to 5 bodies, and that for 6 or more bodies the configuration can never be stable. This places an ideal limit of 5:1 on the aspect ratio of a rubble pile body's shape. 

\end{abstract}

\section{Introduction}
\label{intro}

When gravitating bodies are modeled as rigid bodies with finite density, their gravitational dynamics becomes more rich and can include many additional stable and unstable relative equilibria that do not exist in the classical point-mass $N$ body problem. The study of the gravitational interaction between two rigid bodies is quite classical, tracing back to Lagrange \cite{lagrange1780theorie}, and more recently reconsidered by a wide range of authors from the mathematical, scientific and engineering community. A partial sampling includes Duboshin \cite{duboshin1958differential}, Wang et al. \cite{wang}, Maciejewski \cite{maciejewski}, Beck and Hall \cite{beck1998relative}, Scheeres \cite{scheeres_F2BP} and Moeckel \cite{moeckel2017minimal}, to give an incomplete list. 
More recently the problem has been reconsidered, motivated in part by the discovery that small asteroids are often best modeled as a collection of self-gravitating, rigid components resting on each other and rotating \cite{fujiwara_science}. This generalizes the study of interactions between $N$ rigid bodies to include cases where the bodies can be in physical contact, and have been studied in a series of papers by Scheeres \cite{scheeres_reconfiguration,scheeres_minE,F3BP_scheeres,F4BP_chapter}.  
Key results from these studies include the identification of minimum energy configurations at any level of angular momentum, usually with components resting on each other, and conditions under which these systems can undergo dramatic reconfigurations including fission of components. A recent theoretical advancement was a proof by Moeckel that an orbiting system of rigid bodies cannot be energetically stable for $N \ge 3$ \cite{moeckel2017minimal}, implying that minimum energy configurations for $N\ge3$ rigid bodies must consist of at least some of the bodies being at rest on each other. 

This paper performs an in-depth study of one class of these relative equilibria, the case where the bodies are modeled as equal-sized spheres which rest on each other in a straight-line, reminiscent of the orbital version of the $N$-body Euler solutions from classical celestial mechanics \cite{moulton1920periodic}. Here the questions are different, and are specifically focused on what the minimum spin rate of this configuration is in order to stabilize the system, and what the maximum spin rate of the configuration is when gravitational forces can no longer hold it together and it fissions. For the case of equal masses this problem has been solved for $N=3$ and $N=4$ \cite{scheeres_minE,F4BP_chapter}, and for the case of unequal sized bodies it has been solved for $N=3$ \cite{F3BP_scheeres}, however it has not been explored at higher numbers of components. A specific question of interest is whether there is a maximum number of bodies beyond which this configuration can never be stable. That question is answered in the affirmative in this paper, with the answer being 5. 

While the study of this system is somewhat abstract, the results are instructive in understanding the limits of a natural system to arrange itself in a stable configuration. In particular, it shows that a rubble pile body can never achieve an aspect ratio beyond 5:1, at least when it consists of equal-sized, spherical components. This places an important limit on this otherwise general problem, and motivates variations of it to be studied that involve non-equal sized bodies or non-spherical bodies. 

This paper is organized as follows. First, the general problem statement is given along with the necessary theory for its study. Then the Euler Resting configuration is defined along with its degrees of freedom. Following this the conditions for this system to be in equilibrium and exist are established. Finally, the stability of the configuration is explored and compared with the existence conditions to find the limiting constraint on the number of bodies. Following this, the stability of all equal mass Euler Resting configurations are reviewed. 

\section{Problem Statement}
\label{sec:1}

Consider a system of $N$ spherical bodies of equal diameter $D$ and common density $\rho$. The mass of each body is then ${\cal M} = (\pi/6) \rho D^3$ and their moment of inertia $I = {\cal M} D^2/10$. We will only consider planar configurations of this system, and thus relative to an absolute frame each body has 3 degrees of freedom, two that define its location in the plane and one that rotates the body about an axis perpendicular to the plane. Thus, the degrees of freedom of the system relative to an absolute frame is $3N$. 

\subsection{The Fundamental Functions}

The potential energy, total (planar) moment of inertia about the system center of mass and angular momentum of the system is 
\begin{eqnarray}
	{\cal U} & = & - {\cal G} {\cal M}^2 \sum_{i<j} \frac{1}{d_{ij}} \\
	I_P & = & N  \frac{1}{10} {\cal M} D^2 + \frac{m}{N} \sum_{i<j} d_{ij}^2 \\
	\bfm{H} & = & \sum_{i=1}^N I \omega_i + \frac{{\cal M}}{N} \sum_{i<j} \bfm{d}_{ij}\times\dot{\bfm{d}}_{ij}
\end{eqnarray}
where $\bfm{d}_{ij}$ denotes the relative position vector between two bodies with constraint on the magnitude $d_{ij} \ge D$, and $\omega_i$ is the spin rate of each body. Due to the symmetry of spherical bodies the gravitational potential does not depend on their relative rotational orientation. 

Given these quantities, we can define the amended potential of the (planar) system as
\begin{eqnarray}
	{\cal E} & = & \frac{H^2}{2 I_P} + {\cal U}
\end{eqnarray}
where $H$ is the magnitude of the total angular momentum of the system, which is treated as a constant parameter. Note that the amended potential is only a function of the relative positions of the bodies, and thus is independent of an absolute frame. 

The existence of relative equilibria and their energetic stability for this system can be completely studied through analysis of the amended potential, and is outlined in general in \cite{smaleI,smaleII}, and more specifically in \cite{scheeres_minE,F3BP_scheeres} for the finite density case where bodies can rest on each other. A key aspect of this problem is that minimum energy configurations exist at any level of angular momentum, but may consist of bodies resting on each other. Our current focus is finding the level of angular momentum for when these resting configurations exist and are stable. 

The conditions for a system to be in equilibrium can be established through study of the amended potential  \cite{scheeres_minE}. For the constrained degrees of freedom in this system, denoted as those degrees of freedom at a specific limit $d = d^*$ and only allowed one sided deviations $\delta d \ge 0$, the condition for that degree of freedom to be in equilibrium and to be stable is that $\left.\partial{\cal E}/\partial d\right|_* \ge 0$. For unconstrained degrees of freedom, denoted as $\theta_i$, the condition for a given configuration of the system to be in equilibrium is that $\left.\partial{\cal E}/\partial\theta_i\right|_* = 0$ for all of the unconstrained degrees of freedom evaluated at specific values $\theta_j^*$. Stability in this case is achieved when the Hessian of the amended potential with respect to the unconstrained degrees of freedom is positive definite when evaluated at the equilibrium configuration, $[\partial^2{\cal E}/\partial\theta_i\partial\theta_j]_* > 0$. 

\subsection{The Euler Resting Configuration} 

\begin{figure}
  \includegraphics[width=1.\textwidth]{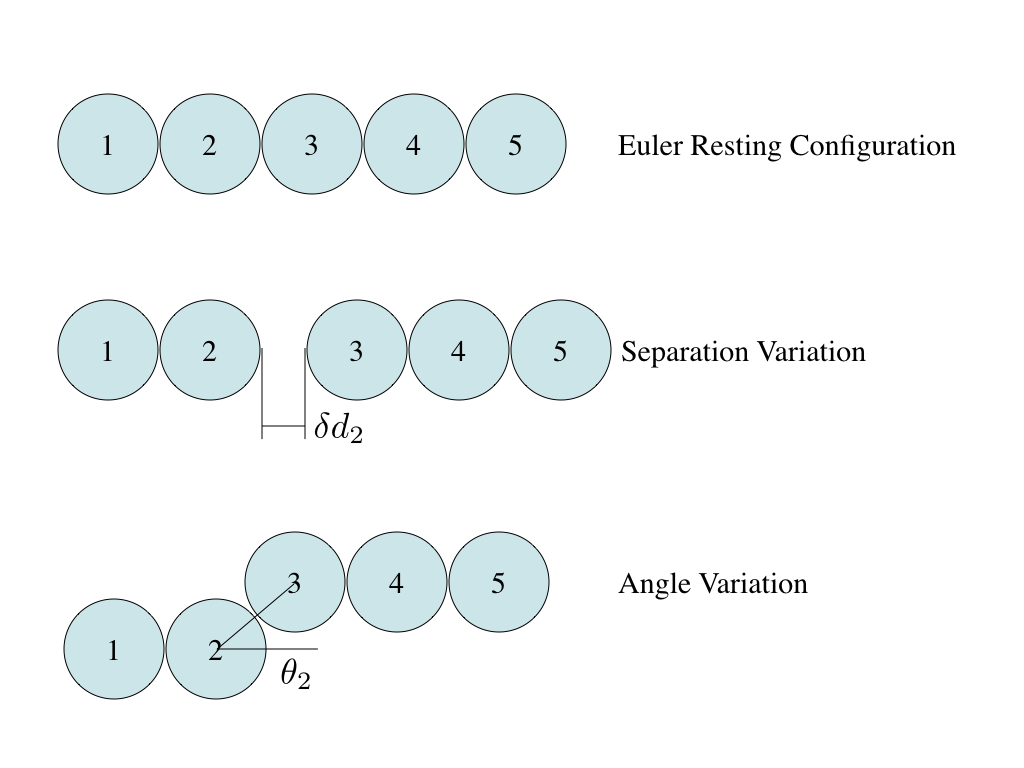}
\caption{Geometry of the Euler Resting configuration and the variations in its degrees of freedom. }
\label{fig:definition}       
\end{figure}

Now define the nominal configuration for this system, shown in Fig.\ \ref{fig:definition}. It is instructive to first review the degrees of freedom of the system under various constraints. The full system has $3N$ degrees of freedom, planar position and rotation for each body. Moving to a relative frame removes the absolute location of the system and the absolute rotation of the system, a total of $3$ degrees of freedom, leaving $3(N-1)$ degrees of freedom. A convenient way to imagine this is to take one body as a reference with a frame embedded in it. Then each of the other $N-1$ body's configuration relative to this one body are specified by their relative location and relative orientation to the body 1 frame, giving the $3(N-1)$ additional degrees of freedom. 

Next, assume that every body is in contact with another body. This is a non-trivial step, as for $N$ bodies there are many different separate topologies for how these components can be attached to each other. For a useful example in the $4$ body problem see \cite{F4BP_chapter}. The current paper only looks at a very specific topology, which has each body in contact with two other bodies, except for the two bodies at the ``ends'' of the chain. Viewing this as a sequential process, starting with body 1, the first constraint is that body 2 rests on it, and then body 3 rests on body 2, etc., yielding $N-1$ constraints on the relative distances between our bodies. 
This reduces the degrees of freedom by $N-1$, leaving $2(N-1)$ degrees of freedom. This can again be imagined, starting with the body 1 frame. Body 2 then has two degrees of freedom of orientation to body 1, its angle of contact relative to body 1, and the rotation of body 2 relative to body 1. Then, body 3 has the same two degrees of freedom relative to body 2, etc. It is important to note that not all relative degrees of freedom are possible for the $i$th body attached to body $(i-1)$, in particular the angle between body $i-2$, $i-1$ and $i$ can never be less than $60^\circ$. However with this nominal topology one can always find an open set of relative orientations such that there are no local constraints on the angular placement of the bodies. 

The final reduction is the no-slip assumption between bodies in contact. This ties the angular location of body $i$ relative to body $i-1$ to the orientation of body $i$ to body $i-1$. Again, an initial ``topology'' must be assumed, which gives the initial orientation of one body relative to another, however once this is chosen then there is only one degree of freedom orienting each sphere relative to the previous one. This adds an additional $N-1$ constraints, yielding the final system with $N-1$ degrees of freedom, when in contact. 

For the nominal configuration, the bodies all rest on a neighbor and their centers of mass all lie along a line. Thus the distance between any two bodies equals the sum of their radii plus the diameters between them. Assume that the bodies are numbered in order, increasing to the right, such that body 1 only has contact on the right by body 2, but that body 2 has contact on both sides (1 and 3), etc., up to body $N$ which only has a contact on the left ($N-1$) and no contact on the right. Then given two bodies $i$ and $j$ chosen such that $i < j$, the distance between them equals
\begin{eqnarray}
	d_{ij} = (j-i) D 
\end{eqnarray}
Thus, when in this configuration the potential energy and moment of inertia are
\begin{eqnarray}
	{\cal U}^* & = & - \frac{{\cal G} {\cal M}^2}{D} \sum_{i<j} \frac{1}{j-i}  \\
	I_P^* & = & \frac{1}{10} N {\cal M} D^2 + \frac{{\cal M} D^2}{N} \sum_{i<j} (j-i)^2
\end{eqnarray}
where the $()^*$ signifies that the system is in the nominal Euler Resting configuration. 

\subsection{Variations About the Euler Resting Configuration}

There are two different types of variations from the nominal configuration that must be considered. These are shown in Fig.\ \ref{fig:definition} and consist of an introduced separation between each body and the rolling of each body on its neighbor. 

The first variation assumes the straight line configuration and introduces positive variations in the distance between any two grains otherwise in contact. As there are $N-1$ contacts between grains, this will consist of $N-1$ possible variations in the state of the system. Due to symmetry this can be reduced in half for an even number of bodies, and by (N-1)/2 for an odd number, as the separation of a set of one bodies to the left is equivalent to the separation of an equal number of bodies to the right. 
In terms of notation, denote the separation between bodies $i$ and $i+1$ to be $d_{i,i+1} = D + d_i$, $i = 1, N-1$ and $d_i \ge 0$. 

The second variation will maintain contact between the bodies but allow them to roll relative to each other, assuming the no-slip condition. While this apparently requires $N-1$ degrees of freedom to fully specify, the symmetry of the problem allows one of these rotational degrees of freedom to be expressed as a combination of all of the others, reducing the number to $N-2$. This can be understood simply with a few examples. First, for $N=2$ the formula states that there are 2 degrees of freedom, however within the amended potential the rolling motion of the two bodies does not appear given the symmetry of the masses. Thus there is only a single degree of freedom, which is the separation between the two spheres. For $N=3$ there are 4 degrees of freedom, however under symmetry there are only 3 degrees of freedom that are apparent, the separation between bodies 1 and 2 ($d_{12}$), 2 and 3 ($d_{23}$), and the angle between body 1 and 3. Then for every additional body just add 2 more degrees of freedom, the distance between $N-1$ and $N$ ($d_{N-1,N}$) and the angle measured by the deflection of body $N$ relative to the original placement of bodies $N-2$ and $N-1$. 
In general, denote the angle that body $i+1$ makes by rolling on body $i$ as $\theta_i$, $i = 1, N-1$ (see Fig.\ \ref{fig:definition} for the specific geometry).  

When the bodies are in contact, the above definitions can develop the general distance between any two bodies, $j$ and $i$, assuming $i<j$, by temporarily introducing a vector system such that the $x$-axis is along the nominal straight line configuration defined by the centers of mass of the bodies, and the $y$-axis is perpendicular to this. Then the coordinates of a body $j$ relative to body 1 in this frame (assuming contact between all neighboring bodies) is 
\begin{eqnarray}
	\bfm{d}_{1j} & = & D \left[ \begin{array}{c}
			\sum_{k=1}^{j-1} \cos\theta_{k} \hat{\bfm{x}} \\
			\\
			\sum_{k=1}^{j-1} \sin\theta_k \hat{\bfm{y}} 
		\end{array} \right] 
\end{eqnarray}
and the relative position between two bodies $j$ and $i$ with $i<j$ is found by subtracting the previous formula, $\bfm{d}_{ij} = \bfm{d}_{1j} - \bfm{d}_{1i}$, to find
\begin{eqnarray}
	\bfm{d}_{ij} & = & D \left[ \begin{array}{c}
			\sum_{k=i}^{j-1} \cos\theta_{k} \hat{\bfm{x}} \\
			\\
			\sum_{k=i}^{j-1} \sin\theta_k \hat{\bfm{y}} 
		\end{array} \right] 
\end{eqnarray}
Finally the relative distance is found by taking the magnitude, which also removes the explicit coordinate system dependance, 
\begin{eqnarray}
	d_{ij}^2 & = & D^2 \left[ \left( \sum_{k=i}^{j-1} \cos\theta_{k} \right)^2 + \left( \sum_{k=i}^{j-1} \sin\theta_{k} \right)^2 \right] 
\end{eqnarray}
It is easy to verify that these distances reduce to the nominal case when $\theta_k = 0$, or indeed whenever all of the angles equal the same value. 

\section{Equilibrium Conditions}

Having the Euler Resting configuration properly specified, the conditions under which this configuration can exist as an equilibrium can be explored. First consider the simpler case of showing that the system as stated will be in equilibrium when $\theta_i = 0$, assuming contact between all components. Then the conditions for when the system can rest on each other and thus be in equilibrium will be considered. 

\subsection{Angular Equilibrium Conditions}

Under the assumption that all of the bodies are resting on each other (to be reviewed in the following subsection), it can be shown that the condition when all of the bodies are aligned is in fact an equilibrium condition. To do this one must show that the gradient of the amended potential with respect to each angular degree of freedom is in fact equal to zero when evaluated at the nominal, or $\left. \partial{\cal E}/\partial\theta_m\right|_{\theta_n=0} = 0$ for all $m$ and $n$. 

Evaluating this condition in the amended potential yields
\begin{eqnarray}
	\frac{\partial {\cal E}}{\partial \theta_m} & = & - \frac{H^2}{2 I_P^2} \frac{\partial I_P}{\partial\theta_m} + \frac{\partial {\cal U}}{\partial\theta_m} 
	\label{eq:E_partial} \\
	\frac{\partial I_P}{\partial\theta_m} & = & \frac{{\cal M}}{N} \sum_{i<j} \frac{\partial d_{ij}^2}{\partial\theta_m} 
	\label{eq:IP_partial} \\
	\frac{\partial {\cal U}}{\partial\theta_m} & = & \frac{1}{2}{\cal G}{\cal M}^2 \sum_{i<j} \frac{1}{d_{ij}^3} \frac{\partial d_{ij}^2}{\partial\theta_m}
	\label{eq:U_partial} 
\end{eqnarray}
Thus, the key component is the gradient of $d_{ij}^2$ with respect to a single angle, $\theta_m$.  
\begin{eqnarray}
	\frac{\partial d_{ij}^2}{\partial \theta_m} & = & \left\{ \begin{array}{lc}
			0 & m \le i-1 \\
			2 D^2 \left[ - \sin\theta_m \sum_{k=i}^{j-1}\cos\theta_k + \cos\theta_m \sum_{k=i}^{j-1}\sin\theta_k  \right] \ & 
				\ i \le m \le j-1 \\
			 0 & m \ge j 
					\end{array} \right. 
\end{eqnarray}

Given this, it is simple to note that when substituting in the value of zero for all nominal angles $\theta_k$ or $\theta_m$ that the gradient is zero. Thus, the Euler Resting configuration is in equilibrium for the case when $\theta_n = 0$. It is interesting to note that the equilibrium condition is also satisfied if $\theta_n = \theta$, i.e., when each angle is at the same value. Then the two terms in the bracket are equal and opposite and cancel each other. This, in fact, represents the additional symmetry in this system noted earlier, which can be removed by requiring one of the angles to be fixed at zero. 

\subsection{Resting Equilibrium Conditions}

The equilibrium condition for a constrained degree of freedom $d_i = 0$ chosen such that $\delta d_i \ge 0$ is that $\left.\partial{\cal E}/\partial d_i\right|_* \ge 0$. To evaluate this condition assume that the system is in its unconstrained equilibrium with $\theta_n = 0$ and that each of the distance variations is tested in turn. 

To analyze the constrained degree of freedom for the contact between any two neighboring bodies, consider the effect of an increase between the distance between two neighboring bodies, $m$ and $m+1$, defined earlier as $d_{m}$. Given this deviation, the general distance between two particles $i$ and $j$, with $i < j$, will be denoted as $(j-i) D + \delta_m d_{i,j}$ where
\begin{eqnarray}
	\delta_{m} d_{i,j} & = & \left\{ \begin{array}{lc}
			0  & m-1 \le i \\
			d_{m} \ & \ 
			i \le m \le j-1 \\
			0 & m \ge j 
				\end{array} \right. 
\end{eqnarray}
More generally, the partial of $d_{ij}$ with respect to the distance $d_{m}$ is
\begin{eqnarray}
	\frac{\partial d_{ij}}{\partial d_{m}} & = & \left\{ \begin{array}{lc}
			0  & m \le i-1 \\
			1 \ & \ 
			i \le m \le j-1 \\
			0 & m \ge j
				\end{array} \right. 
\end{eqnarray}

Take the partial of the amended potential with respect to $d_m$ and evaluate it at $d_i = 0$ to find
\begin{eqnarray}
	\left. \frac{\partial {\cal E}}{\partial d_{m}} \right|_* & = & - \frac{H^2}{2 I_P^2} \left. \frac{\partial I_P}{\partial d_{m}}\right|_*  + \left. \frac{\partial {\cal U}}{\partial d_{m}}\right|_*  \\
	\frac{\partial I_P}{\partial d_{m}} & = & \frac{2 {\cal M}}{N} \sum_{i<j} d_{ij} \left. \frac{\partial d_{ij}}{\partial d_{m}}\right|_*  \\
	\frac{\partial {\cal U}}{\partial d_{m}} & = & {\cal G}{\cal M}^2 \sum_{i<j} \frac{1}{d_{ij}^2} \left. \frac{\partial d_{ij}}{\partial d_{m}}\right|_* 
\end{eqnarray}
When evaluated at the nominal resting position the value of $d_{ij} = D (j-i)$. Thus the general summation will be of the form
\begin{eqnarray}
	\sum_{i<j} f(d_{ij}) \frac{\partial d_{ij}}{\partial d_{m}}  & = & \sum_{i=1}^{N-1} \sum_{j=i+1}^N f(j-i) \frac{\partial d_{ij}}{\partial d_{m}} \\
	& = & \sum_{i=1}^{m} \sum_{j=m+1}^{N} f(j-i) \\
	& = & \sum_{k=1}^{N-1} f(k) \min[ k, N-k, m, N-m ]
\end{eqnarray}

There are two separate summations which must be evaluated, one of which can be done in closed form and the other must in general be computed numerically. The gradient of the moment of inertia yields $f(d_{ij}) = D(j-i)$ and the summation 
\begin{eqnarray}
	\frac{\partial I_P}{\partial d_{m}} & = & \frac{2 {\cal M} D}{N} \sum_{k=1}^{N-1} k \min[ k, N-k, m, N-m ] \\
	& = &  {\cal M} D m (N-m) 
\end{eqnarray}
The gradient of the gravitational potential yields $f(d_{ij}) = \frac{1}{D^2 (j-i)^2}$ and the summation 
\begin{eqnarray}
	\frac{\partial {\cal U}}{\partial d_{m}} & = & \frac{{\cal G}{\cal M}^2}{D^2} \sum_{k=1}^{N-1} \frac{1}{k^2} \min[ k, N-k, m, N-m ]
\end{eqnarray}
which does not have a closed form and must be summed using an algorithm. The general form of this summation can be denoted as
$g_s(m,N) = \sum_{k=1}^{N-1} k^{-s} \min[ k, N-k, m, N-m ]$, where for the above the function is $g_2(m,N)$. Note that the function can be solved in closed form when $g_{-s}(m,N)$, such as for the moment of inertia term, and that $g_s(1,\infty) = \zeta(s)$, the Riemann zeta function. Finally, note the symmetry $g_s(m,N) = g_s(N-m,N)$.

With these results, the original condition can now be explicitly expressed as
\begin{eqnarray}
	\left. \frac{\partial {\cal E}}{\partial d_{m}}\right|_* & = & 
	- \frac{H^2}{2 I_P^2} {\cal M} D m (N-m) +  \frac{{\cal G}{\cal M}^2}{D^2} g_2(m,N) 
\end{eqnarray}
and the condition $\left. \partial{\cal E}/\partial d_{m}\right|_* \ge 0$ yields a constraint on the angular momentum of the system for the equilibrium to exist
\begin{eqnarray}
	\left( \frac{H}{I_P}\right)^2 \frac{D^3}{{\cal G}{\cal M}} \le \frac{2}{m(N-m)} g_2(m,N)
\end{eqnarray}
for $m = 1, 2, \ldots, N-1$. However, given the symmetry obtained in replacing $m$ by $N-m$ one only needs consider $m=1, 2, \ldots, N/2$ for $N$ even and up to $(N-1)/2$ for $N$ odd. 

Note that the ratio $H/I_P$ equals the total spin rate of the relative equilibrium, thus this can be interpreted directly as a constraint on the total spin rate as well. The expression tells us the upper limit on the system spin rate before the system will separate between the body $m-1$ and $m$. Since this condition must hold for all pairs of bodies, the final constraint is the minimum of this limit for all values of $m$. For the leading term, it is simple to note that that choosing $m \sim N/2$ will in general minimize the leading portion, however the $g_2$ function is maximized at this point. Combining the two expressions we find that the minimum occurs at $m = N/2$ for $N$ even or $m = (N\pm1)/2$ for $N$ odd.  This then defines the spin rate at which the Euler Resting configuration will fission, and we note that this occurs by the body splitting ``in half''. This can be predicted from first principles analysis as well, as the stress from centrifugal accelerations will be the highest at the mid-point of the spinning system and hence will be the first point at which it becomes positive and the system separates. 

Thus the fission spin limit for the Euler Resting configuration is
\begin{eqnarray}
	\bar{\Omega}^2_F & = & \left\{ 
		\begin{array} {lr} 
			\frac{2}{P^2} g_2(P,2P) & \ N = 2P \\
			& \\
			\frac{2}{P(P+1)} g_2(P,2P+1) & \ N = 2P+1 
		\end{array} \right.  			
		\label{eq:fission}
\end{eqnarray}
where $\bar{\Omega}$ denotes a non-dimensional spin rate of the system, $H/I_P$, normalized by $\sqrt{{\cal G}{\cal M}/D^3}$. 

\section{Stability of the Euler Resting Configuration}

The previous analysis establishes the existence of the Euler Resting configuration for a particular alignment and for spin rates less than some value. For the constrained degrees of freedom, they are stable whenever they exist, and thus need not be analyzed further. The same is not true for the unconstrained equilibria defined by $\theta_n = 0$. The condition for these equilibria to be stable is that the Hessian of the amended potential evaluated at the equilibrium position be positive definite, or that the matrix $[ \left. \partial^2{\cal E} / \partial\theta_m \partial\theta_n \right|_*] > 0$. 

\subsection{Evaluating the Hessian}

Starting from Eqn.\ \ref{eq:E_partial} one finds
\begin{eqnarray}
	\left. \frac{\partial^2 {\cal E}}{\partial \theta_m\partial\theta_n}\right|_* & = & - \frac{H^2}{2 I_P^2} \left. \frac{\partial^2 I_P}{\partial\theta_m\partial\theta_n} \right|_*
		+ \frac{H^2}{I_P^3} \left. \frac{\partial I_P}{\partial\theta_m}\right|_* \left. \frac{\partial I_P}{\partial\theta_n} \right|_*
		+ \left. \frac{\partial^2 {\cal U}}{\partial\theta_m\partial\theta_n} \right|_*
\end{eqnarray}
however evaluating this at the equilibrium condition removes the second term as $\left. \frac{\partial I_P}{\partial\theta_m}\right|_* =0$. Evaluating the second partials of the moment of inertia and gravitational potential yields 
\begin{eqnarray}
	\left. \frac{\partial^2 I_P}{\partial\theta_m\partial\theta_n}\right|_* & = & \frac{{\cal M}}{N} \sum_{i<j} \left. \frac{\partial^2 d_{ij}^2}{\partial\theta_m\partial\theta_n}\right|_* \\
	\left. \frac{\partial^2 {\cal U}}{\partial\theta_m\partial\theta_n}\right|_* & = & \frac{1}{2}{\cal G}{\cal M}^2 \sum_{i<j}\left.  \frac{1}{d_{ij}^3} \frac{\partial^2 d_{ij}^2}{\partial\theta_m\partial\theta_n}\right|_*
\end{eqnarray}
where for the potential energy term recall that the first partial of $d_{ij}^2$ is zero when evaluated at the equilibrium, leaving only the term involving the second partial. 

Then, assuming that $m\le n$, the second partial is 
\begin{eqnarray}
	\frac{\partial^2 d_{ij}^2}{\partial \theta_m\partial\theta_n} & = & \left\{ \begin{array}{lc}
			0 & m \le i-1 \\
			2 D^2 \left[ 1 - \sum_{k=i}^{j-1}\left[ \cos\theta_k\cos\theta_m + \sin\theta_k\sin\theta_m  \right] \right] \ & 
				\ i \le m = n \le j-1  \\
			2 D^2 \left[ \sin\theta_m \sin\theta_n + \cos\theta_m \cos\theta_n  \right] \ & 
				\ i \le m < n \le j-1\\
			 0 & n \ge j 
					\end{array} \right. 
\end{eqnarray}
and evaluating this at the equilibrium condition yields
\begin{eqnarray}
	\left. \frac{\partial^2 d_{ij}^2}{\partial \theta_m\partial\theta_n}\right|_* & = & 2 D^2 \left\{ \begin{array}{lc}
			0 & m \le i-1 \\
			 \left[ 1 - (j-i) \right] \ & 
				\ i \le m = n \le j-1 \\
			1 \ & 
				\ i \le m < n \le j-1 \\
			 0 & n \ge j 
					\end{array} \right. 
\end{eqnarray}

The second partial of the moment of inertia is then 
\begin{eqnarray}
	\left. \frac{\partial^2 I_P}{\partial\theta_m\partial\theta_n}\right|_* & = & \frac{{\cal M}}{N} \sum_{i<j} \frac{\partial^2 d_{ij}^2}{\partial\theta_m\partial\theta_n} {\partial\theta_m\partial\theta_n} \\
	& = & \frac{2{\cal M} D^2}{N} \sum_{i=1}^m \sum_{j=n+1}^N 
		\left\{ \begin{array}{lr}
			1 - (j - i) & m = n \\
			1 & m < n
							\end{array} \right. \\
	& = & \frac{2{\cal M} D^2}{N}  
		\left\{ \begin{array}{lr}
			- \frac{1}{2} m (N-m) (N-2) & m = n \\
			m(N-n) & m < n
							\end{array} \right. 
\end{eqnarray}
The second partial of the potential energy is then
\begin{eqnarray}
	\left. \frac{\partial^2 {\cal U}}{\partial\theta_m\partial\theta_n}\right|_* & = & \frac{{\cal G}{\cal M}^2}{2D^3} \sum_{i<j} \frac{1}{(j-i)^3} \frac{\partial^2 d_{ij}^2}{\partial\theta_m\partial\theta_n} {\partial\theta_m\partial\theta_n} \\
	& = & \frac{{\cal G}{\cal M}^2}{D} \sum_{i=1}^m \sum_{j=n+1}^N 
		\left\{ \begin{array}{lr}
			\frac{1 - (j - i)}{(j-i)^3} & m = n \\
			\frac{1}{(j-i)^3} & m < n
							\end{array} \right. 
\end{eqnarray}
Introduce a generalization of the previous summation formula
\begin{eqnarray}
	\sum_{i=1}^m \sum_{j=n+1}^N f(j-i) & = & \sum_{k=n-m+1}^{N-1} f(k) \min[ k-(m-n), N-k, m, N-n ]
\end{eqnarray}
and the generalized summation function
\begin{eqnarray}
	h_s(m,n,N) & = & \sum_{k=n-m+1}^{N-1} \frac{1}{k^s} \min[ k-(n-m), N-k, m, N-n ]
\end{eqnarray}
where $m\le n$ and $h_s(m,m,N) = g_s(m,N)$. 
Then the potential energy partials can be expressed as
\begin{eqnarray}
	\left. \frac{\partial^2 {\cal U}}{\partial\theta_m\partial\theta_n}\right|_* & = & \frac{{\cal G}{\cal M}^2}{D}  
		\left\{ \begin{array}{lr}
			g_3(m,N) - g_2(m,N)  & m = n \\
			h_3(m,n,N) & m < n
							\end{array} \right. 
\end{eqnarray}

Thus, there are two possible forms for the second partial of the amended potential, denoted in short hand as ${\cal E}_{mn}$. 
\begin{eqnarray}
	{\cal E}_{mm} & = & \frac{H^2}{2 I_P^2} {{\cal M} D^2} \frac{m (N-m) (N-2)}{N}   
		- \frac{{\cal G}{\cal M}^2}{D} \left[ g_2(m,N) - g_3(m,N) \right] 
\end{eqnarray}
where $g_2 > g_3$ in general. 
For the off-diagonal terms ${\cal E}_{mn} = {\cal E}_{nm}$ and, for $m < n$
\begin{eqnarray}
	{\cal E}_{mn} & = & - \frac{H^2}{I_P^2} {\cal M} D^2\frac{m(N-n)}{N}  
		+ \frac{{\cal G}{\cal M}^2}{D} h_3(m,n,N) 
\end{eqnarray}

The condition for the system to be stable is that the matrix $[{\cal E}_{ij}]$ be positive definite. There are a number of ways to set up the necessary and sufficient conditions. Two that we use here are to determine that the principal minors are all positive, or that the eigenvalues of the matrix are all positive. To aid in these discussions rewrite the Hessian in a scaled version where $\bar{\cal E} = {\cal E} / ({\cal G}{\cal M}^2/D)$ and $\bar{\Omega} = (H/I_P) / \sqrt{{\cal G}{\cal M}/D^3}$. Then 
\begin{eqnarray}
	\bar{\cal E}_{mm} & = & \bar{\Omega}^2 \frac{m (N-m) (N-2)}{2 N}   
		-  \left[ g_2(m,N) - g_3(m,N) \right] \\
	\bar{\cal E}_{mn} & = & - \bar{\Omega}^2 \frac{m(N-n)}{N}  
		+  h_3(m,n,N) 
\end{eqnarray}

\subsection{Necessary Conditions}

A useful necessary condition is that all of the diagonals of the matrix be positive, ${\cal E}_{mm} > 0$, as this is a precondition for the principal minors to be positive. This yields a simple condition on the normalized spin rate of the system, 
\begin{eqnarray}
	\bar{\Omega}^2 & > & \frac{2N}{m(N-m)(N-2)} \left[ g_2(m,N) - g_3(m,N) \right] \label{eq:nec}
\end{eqnarray}
with the additional restriction that $N \ge 3$, and that $1\le m \le N-1$. From this it is clear that $H^2$ can always be chosen small enough so that the Euler Resting configuration is definitely unstable. Similarly, the angular momentum must be greater than the maximum of the right-hand side for stability to be a possibility. 

This condition is in direct competition for the existence condition for the Euler Resting configuration, summarized in Eqn.\ \ref{eq:fission}. 
Comparing Eqn.\ \ref{eq:nec} with Eqn.\ \ref{eq:fission} one finds that the angular momentum for the necessary condition to be satisfied exceeds the existence condition limit for $N \ge 9$. Thus this sets a limit on the size of an Euler Resting configuration, requiring a more detailed analysis only up to this value. Of course, sharper necessary conditions can be tested, such as the trace of the matrix, however these become progressively more complex to analyze, making the jump directly to the sufficiency conditions reasonable. 

\subsection{Sufficient Conditions}

To check the sufficient condition, the positive definiteness of the amended potential Hessian must be checked across a range of angular momenta between the necessary condition for stability up to the fission limit. To make this computation definitive, the sufficiency condition can be evaluated at the fission limit value given in Eqn.\ \ref{eq:fission}, to check whether the stability condition is satisfied up to this level. Carrying out the computation shows that for $N=5$ the sufficiency condition is definitely satisfied when the system is evaluated at the fission condition, however at $N=6$ and higher it is violated with at least one negative eigenvalue. 

 Note that the case $N=2$ is trivially stable when it exists, and that the cases $N=3,4$ have been studied in detail previously. Thus, in the current paper a complete analysis is only carried out for the case $N=5$, which is also the only remaining case to be studied that can be stable. 
 
 \subsection{Stability limits of the $N=5$ case}
 
 Per the earlier discussion on the number of free degrees of freedom, only the variation of $5-2 = 3$ degrees of freedom need to be considered, which can be arbitrarily chosen out of the set $\theta_i, i=1,2,3,4$. For notational ease we consider the first three, giving the specific entries for the matrix as follows:
\begin{eqnarray}
	\bar{\cal E}_{11} & = & \bar{\Omega}^2 \frac{6}{5}   
		-  \left[ g_2(1,5) - g_3(1,5) \right] \\
	\bar{\cal E}_{22} & = & \bar{\Omega}^2 \frac{9}{5}   
		-  \left[ g_2(2,5) - g_3(2,5) \right] \\
	\bar{\cal E}_{33} & = & \bar{\Omega}^2 \frac{9}{5}   
		-  \left[ g_2(3,5) - g_3(3,5) \right] \\
	\bar{\cal E}_{12} & = & - \bar{\Omega}^2 \frac{3}{5}  
		+  h_3(1,2,5) \\
	\bar{\cal E}_{13} & = & - \bar{\Omega}^2 \frac{2}{5}  
		+  h_3(1,3,5) \\
	\bar{\cal E}_{23} & = & - \bar{\Omega}^2 \frac{4}{5}  
		+  h_3(2,3,5) 
\end{eqnarray}
Where the values of the $g_s$ and $h_s$ functions are given in Table \ref{tab:functions}. 

\begin{table}[h!]
\centering
\caption{Specific values of the functions $g_s$ and $h_s$ for $N=5$.}
\label{tab:functions}
\begin{tabular}{| c || r | }
\hline
$g_2(1,5)$ & 1.42361111193895 \\
$g_2(2,5)$ & 1.78472222387791 \\
$g_2(3,5)$ & 1.78472222387791 \\
$g_3(1,5)$ & 1.17766203731298 \\
$g_3(2,5)$ & 1.33969907462597 \\
$g_3(3,5)$ & 1.33969907462597 \\
$h_3(1,2,5)$ & 0.17766203731298 \\
$h_3(1,3,5)$ & 0.05266203731298 \\ 
$h_3(2,3,5)$ & 0.21469907462597 \\
\hline
\end{tabular}
\end{table}

The range of values to be explored go up to $\bar{\Omega}_F^2 = g_2(2,5) / 3$ or $\bar{\Omega}_F = 0.7713024\ldots$, which is the fission spin rate for $N=5$. 
Through computation we find that the determinant is the defining condition for the stability of this system, which is to be expected as it is the product of the eigenvalues. Evaluating the determinant as the controlling condition, we have a cubic equation in $\bar{\Omega}^2$ to be considered. This can be iteratively solved to find the zero crossing, which occurs at a value of $\bar{\Omega}_S / \bar{\Omega}_F = 0.674064\ldots$, or at a normalized value of $\bar{\Omega}_S = 0.5199071\ldots$. Thus, the Euler Resting Configuration for $N=5$ is stable when the normalized spin rate lies in the interval
\begin{eqnarray}
	0.5199071\ldots \le \bar{\Omega} \le 0.7713024\ldots
\end{eqnarray}
At lower values of angular momentum the system is unstable relative to the angular variations. At higher values of angular momentum the system will fission into two components, one with two connected bodies and the other with three. 

\section{Discussion}

It is instructive to compare the different spin limits for the different Euler resting configurations. The comparable spin rate at which the system becomes stabilized, as a fraction of the fission spin rate, is zero for the $N=2$ case, is $\sqrt{0.3} = 0.5477$ for the $N=3$ case, and is $0.6343$ for the $N=4$ case.  Thus, the $N=5$ case has the smallest interval of normalized spin rate for stability, as measured by the fission spin rate. The complete fission and stabilization non-dimensional spin rates as a function of $N$ is given in Table \ref{tab:fission}. 

To bring these results back to the physical domain, the dimensional value of these limiting spin rates will be considered for typical body densities. First, note that the normalizing spin rate is only a function of the density of the bodies, as $\sqrt{ {\cal G}{\cal M}/D^3} = \sqrt{{\cal G}\pi \rho / 6}$, where ${\cal G} = 6.6702\times10^{-11}$ m$^3$/ kg / s$^2$ and where $\rho$ is the body density measured in units of kg/m$^3$. Typical values of density for natural solar system bodies can range from 500 for cometary bodies up to 5000 for boulders with low porosity. This corresponds to a range of normalizing frequencies from $1.3215\times10^{-4} \rightarrow 4.1788\times10^{-4}$ rad/s, or equivalently normalizing rotation periods from $13.2076 \rightarrow 4.1766$ hr. 
\begin{table}[h!]
\centering
\caption{Non-dimensional fission and stabilization and fission spin rates. To account for densities ranging from $500\rightarrow5000$ kg/m$^3$ multiply by $1.3215\times10^{-4} \rightarrow 4.1788\times10^{-4}$ rad/s, respectively.}
\label{tab:fission}
\begin{tabular}{| c || c | c |}
\hline
$N$ & $\bar{\Omega}_F$ & $\bar{\Omega}_S$ \\
\hline
2 &  $1.4142$ & 0 \\
3 & $1.1180$ & 0.6123 \\
4 & 0.8975 & 0.5693 \\
5 & 0.7713 & 0.5199 \\
\hline
\end{tabular}
\end{table}

That the Euler Resting configuration is only stable up to $N=5$ places a limit on the elongation that a collection of resting particles can have at 5:1. While such an elongation in a natural body is extreme, it is interesting to note that the first interstellar object, 1I/`Oumuamua, was found to have an elongation commensurate with this level. A fundamental question of interest for future studies is whether a similar elongation limit holds when systems of more bodies are considered, arranged to have additional components stacked on each other in directions transverse to the long axis. 


%

\begin{acknowledgements}
This research was supported by NASA grant NNX14AL16G. The author declares no conflict of interests. 
\end{acknowledgements}


\bibliographystyle{plain}
\bibliography{../../bibliographies/biblio_article,../../bibliographies/biblio_books,../../bibliographies/biblio_misc}


\end{document}